\documentclass[12pt]{article}
\usepackage{amssymb,amsmath}
\usepackage{caption}
\usepackage[noblocks]{authblk}
\begin{document}
\textwidth 10.0in 
\textheight 9.0in 
\topmargin -0.60in

\title{ The Double Slit Experiment With Polarizers}
\author[1]{M. Holden}
\author[1,2]{D.G.C. McKeon}
\author[3] {T.N. Sherry}
\affil[1] {Department of Applied Mathematics, The
University of Western Ontario, London, ON N6A 5B7, Canada} 
\affil[2] {Department of Mathematics and
Computer Science, Algoma University, Sault St.Marie, ON P6A
2G4, Canada}
\affil[3] {School of Mathematics, Statistics and Computer Science,
  National University of Ireland Galway, University Road, Galway, Ireland}

\maketitle

\maketitle

\begin{abstract}
 The double slit experiment provides a standard way of demonstrating how quantum mechanics works. We consider
 modifying the standard arrangement so that a photon beam incident upon the double slit encounters a polarizer 
 in front of either one or both of the slits. 
\end{abstract}

The role of the wave function in quantum theory is well illustrated by considering the effect of a double slit on a beam 
of either photons or electrons, as is discussed in many texts (see, for example, ref. [1]). This basic experimental 
situation has been modified in several ways in order to probe more deeply into the whole problem of measurement and 
entanglement in quantum physics (for example, see refs. [2-5]). In this short note we consider another way of examining 
how quantum mechanics works that employs a double slit interacting with a beam of photons.

What we propose is placing a linear polarizer in front of either one or both of the double slits. The polarizer permits 
only photons of a given polarization to pass through the slit with which it is associated.

Let us first place a ``vertical'' polarizer in front of slit ``1'' and not place a polarizer in front of slit ``2''. The 
state $|\psi_i>$ of the incoming beams is taken to be 
\begin{equation}
|\psi_i> = \frac{a|\psi^V> + b|\psi^H>}{\sqrt{|a|^2 + |b|^2}}\,, 
\end{equation}
where $|\psi^V>$ and $|\psi^H>$ are two orthonormal basis states referring to vertically and horizontally polarized photons. 
If $a = 1$, $b = 0$ then photons in the incoming beam could go through either of the two slits, while if $a = 0$, $b = 1$ then 
these photons could only pass through slit ``2''. If the beam is in state $|\psi_f>$ after it passes through this device, 
then we see from this that 
\begin{equation}
|\psi_f> = \frac{1}{\sqrt{2}} \left[ \frac{a}{\Lambda} (|\psi^V_1> + |\psi_2^V> )\right] + \frac{b}{\sqrt{2}\Lambda} |\psi^H_2>.
\end{equation}
where $\Lambda = \sqrt{|a|^2 + |b|^2}$ and $|\psi^{I\!\!P}_k>$ refers to a state involving photons that have passed through 
slit ``k'' $(k = 1,2)$ and have polarization $I\!\!P (I\!\!P = V,H)$. (Of course, not all of the incident beam passes through 
the device as slit ``1'' is opaque to horizontally polarized photons which accounts for $|\psi_f>$ not being normalized to 
one.)

From eq. (2) we see that since 
\begin{align}
<\psi_f|\psi_f> &= \frac{1}{2} \frac{|a|^2}{\Lambda^2} \left( <\psi_1^V|\psi_1^V> + <\psi_2^V|\psi_2^V> + 2 Re <\psi_1^V|\psi_2^V> \right)\nonumber \\
 &\qquad + \frac{1}{2} \frac{|b|^2}{\Lambda^2} <\psi_2^H|\psi_2^H> 
\end{align}
the pattern of photons that have passed through this apparatus on a screen is the superposition of a diffraction pattern resulting 
from the vertically polarized photons passing through both slits and the horizontally polarized photons only passing through slit 
``2''; these two patterns (one for each polarization) are superimposed on each other.  If our detector were sensitive to only vertically polarized photons, then we would not know which slit the photons would have passed through and an interference pattern results.  However, if our detector were only triggered by horizontally polarized photons, then the detector would not display an interference pattern as the polarizer in front of slit ``1'' has resulted in all horizontally polarized photons passing through slit ``2''.

We now consider the effect of having not only a vertical polarizer in front of slit ``1'', but also a second polarizer in 
front of slit ``2'' that is inclined at an angle $\theta$ with respect to the vertical so that 
\begin{equation}
|\psi^\theta > = \cos \theta |\psi^V > + \sin \theta |\psi^H> .
\end{equation}
If $|\psi^\theta >$ and $|\psi^{\theta + \pi/2}>$ are a pair of orthonormal polarization states, then 
\begin{equation}
|\psi^{\theta + \pi/2}> = -\sin \theta |\psi^V> + \cos \theta |\psi^H>.
\end{equation}
Thus the state $|\psi_i>$ of eq. (1) can be written as 
\begin{align}
|\psi_i> &= \frac{1}{\Lambda} \left[ (a \cos \theta + b \sin\theta) |\psi^\theta> \right.\\
 & \left. \qquad  + (-a \sin \theta + b \cos\theta) |\psi^{\theta + \pi/2}>\right]\,, \nonumber
\end{align}
and so from (1) and (6), the state of the beam after it has passed through these two polarizers is 
\begin{align}
|\psi_f> &= \frac{1}{\sqrt{2}} \frac{1}{\Lambda} \left[ a |\psi_1^V> + (a \cos\theta + b \sin\theta ) |\psi_2^\theta >\right] \\
&=  \frac{1}{\sqrt{2}} \frac{1}{\Lambda} \left[ a |\psi_1^V> + (a \cos\theta + b \sin\theta ) (\cos\theta |\psi_2^V> + 
\sin\theta |\psi_2^H>)\right].\nonumber
\end{align}
It is again apparent that provided $a \neq 0$, $(a \cos\theta + b\sin\theta)\cos\theta \neq 0$ there will be interference 
between vertically polarized light contributing to $|\psi_1^V>$ and $|\psi_2^V>$. If $a = 0$, $\cos\theta = 0$ or 
$a\cos\theta + b\sin\theta = 0$ then the interference pattern is lost as in these cases, the polarizers allow us to 
determine which slit the beam passes through.  Furthermore, there is superimposed on this pattern those horizontally 
polarized photons that have passed only through slit ``2''. 

We note that since the coefficients of $|\psi_1^V>$ and $|\psi_2^V>$ in eq. (7) are not in general equal, which means that 
the shape of the diffraction pattern arising from these two states is not the same as it would be if the polarizers were 
not present. To see how this comes about, note first that the difference in distance from the two slits to a point $x$ 
units above a point on a screen behind the two slits that is directly opposite the slits is [1]
\begin{align}
\delta &= \ell_2 - \ell_1 \nonumber \\
&= \sqrt{D^2 + \left(x + \frac{d}{2}\right)^2} - \sqrt{D^2 - \left(x - \frac{d}{2}\right)^2} \nonumber \\
&\approx \frac{xd}{D}
\end{align}
where $d$ is the distance between the slits and $D$ is the distance from the slits to the screen. In general the form 
of the waves arriving at $x$ on the screen from the two slits is (provided $\lambda$ is the wave length of the beam)
\begin{align}
A_1 \cos &\left( \frac{2\pi}{\lambda} \ell_1\right) + A_2 \cos \left( \frac{2\pi}{\lambda} (\ell_1 + \delta)\right)\nonumber \\
& \qquad \approx \cos \left( \frac{2\pi}{\lambda} \ell_1\right) \left[A_1 + A_2 \cos \left( \frac{2\pi}{\lambda} \delta\right)
\right]
\end{align}
with $A_1$ and $A_2$ read off of eq. (7). Only if $A_1 = A_2 = A$, does eq. (9) then reduce to the standard form  
\begin{equation}
= 2A\cos \left( \frac{2\pi}{\lambda} \ell_1\right)
 \cos^2 \left( \frac{\pi\delta}{\lambda} \right).\nonumber
\end{equation}

Only if $\theta = \pi/2$ do our pair of polarizers act as ``which way'' detectors. In this case slit ``1'' will only 
allow vertically polarized photons to pass through wile slit ``2'' admits only horizontally polarized photons, so that 
no interference pattern emerges, as can be seen from eq. (7).  For $\theta \neq \pi/2$, then a vertically polarized 
photon can in principle pass through either slit and an interference pattern results; superimposed on this interference 
pattern for vertically polarized photons is a diffraction pattern for horizontally polarized photons which only 
pass through one slit.

This situation can be contrasted with the experiment described in refs. [4-5]. In the set-up described in these references, a device is employed which converts 
linear polarization into circular polarization (either left ``L'' or right ``R'' handed polarizations) without altering 
the beam in any other way.  The device $Q_1$ in front of slit ``1'' converts a beam with $V(H)$ linear polarization into 
$L(R)$ circular polarization while the device $Q_2$ in front of slit ``2'' converts $V(H)$ into $R(L)$. In addition, 
the beam $s$ which is incident upon these slits is entangled with a second beam $p$; if the first beam $s$ has polarization $V(H)$, 
the second beam $p$ has polarization $H(V)$. 

Consequently, since if the polarization of the beam $p$ is measured, we can determine 
the slit through which beam $s$ passed by measuring the circular polarization of the photons that have passed through 
the double slits, and so the presence of $Q_1$ and $Q_2$ serves to eliminate the normal interference pattern resulting from a 
beam passing through the double slits. For example, if beam $p$ has polarization $V$, and if beam $s$ has polarization $L$ after passing through the slits, then we know that beam $s$ has passed through slit ``2'' (ie, $VL \rightarrow 2$; similarly $HL \rightarrow 1$, $VR \rightarrow 1$, $HR \rightarrow 2$), and there is no interference pattern resulting from the beam passing through the slits. 

However, if a device is used to alter the beam $p$ so that its polarization 
cannot be determined, then measuring the circular polarization of the photons emerging from the double slits no longer 
serves to determine which slit the beam passed through and the interference pattern is restored.  It is only by measuring both the polarization of beam $p$ and the circular polarization of beam $s$ after it has passed through the slits that we can infer which slit beam $s$ has passed through. If we lose the ability to measure the polarization of beam $p$ (possibly by passing it through a polarizer at angle $\theta = \pi/4$ to the $V$ and $H$ directions) then we are unable to infer which slit the beam $s$ has passed through and the interference pattern is restored.  Remarkably, eliminating 
the possibility of determining the linear polarization of beam $p$ can occur \underline{after} the circular polarization 
of the beam emerging from the double slits is measured and still the interference pattern is restored. Of course, these two processes (eliminating the possibility of measuring the polarization of $p$ and of measuring the circular polarization of $s$) are separated by a space-like interval; ``after'' refers to the sequence in which events occur in the lab frame but not necessarily in all other inertial frames--causality cannot be violated. 

The devices $Q_1$ and $Q_2$ differs from the polarizers we have considered in this paper when $\theta \neq \pi/2$. By measuring the 
circular polarization of the photons emerging from the double slit and combing this information with knowledge of the  
polarization of beam $s$ as inferred by measuring the polarization of the entangled photons in beam $p$, the  
interference pattern occurring after beam $s$ has passed through the double slit is lost as these measurements together determine which slit a photon has passed through. In contrast, in our 
experimental set up, if $\theta \neq \pi/2$ we cannot determine which slit a vertical photon has passed through, 
even by considering the polarization of a photon entangled with a photon incident on our double slits with polarizers in 
front of each slit\vspace{.2cm}.\\

\noindent
{\Large\bf{Acknowledgments}}\\
We appreciate a discussion with A. Shiekh. R. Macleod had a helpful suggestion.

\end{document}